# Role of dipole-dipole interactions in multiple quantum transitions in magnetic nanoparticles

N. Noginova[1], Yu. Barnakov[1], A. Radocea[2], V.A. Atsarkin[3]

[1] *Center for Materials Research, Norfolk State University, Norfolk, VA*
[2] *Cornell University, Ithaca, NY*
[3] *Kotel'nikov Institute of Radio Engineering and Electronics RAS, Moscow, Russia*

In order to better understand the origin of multiple quantum transitions observed in superparamagnetic nanoparticles, electron magnetic resonance (EMR) studies have been performed on iron oxide nanoparticles assembled inside the anodic alumina membrane. The positions of both the main resonance and "forbidden" (double-quantum, 2Q) transitions observed at the half-field demonstrate the characteristic angular dependence with the line shifts proportional to $3\cos^2\theta-1$, where $\theta$ is the angle between the channel axis and external magnetic field **B**. This result can be attributed to the interparticle dipole-dipole interactions within elongated aggregates inside the channels. The angular dependence of the 2Q intensity is found to be proportional to $\sin^2\theta\cos^2\theta$, that is consistent with the predictions of quantum-mechanical calculations with the account for the mixing of states by non-secular inter-particle dipole-dipole interactions. Good agreement is demonstrated between different kinds of measurements (magnetization curves, line shifts and 2Q intensity), evidencing applicability of the quantum approach to the magnetization dynamics of superparamagnetic objects.

## 1. Introduction

Atomic scale objects are described with the quantum approach while macroscopic systems are the common subject of the classical physics. Nanoscale magnetic objects can be used as a playground to study the transition between the quantum type of behavior and classical picture [1-4]. As was shown earlier [2,3], the electron magnetic resonance (EMR) in magnetic nanoparticles (NP) in certain conditions can be satisfactory described with the "quantatization" approach, considering a nanoparticle of a small size (5-10 nm) as a large exchange-coupled cluster with the spin S ~ $10^2$-$10^3$ of the ground spin multiplet. The EMR spectrum was ascribed to the quantum transitions between the energy levels associated with the projections, *m,* of this "giant spin" onto the direction of the magnetic field (with account made for the magnetic anisotropy), in a way similar to the common electron paramagnetic resonance consideration. Moreover, in a strong resemblance to multiple quantum transitions known for paramagnetic spins, the EMR signal in NPs demonstrates additional low-field signals at the fields $B = B_0/k$ (where $B_0$ is the field of the main resonance and $k = 2, 3, 4...$ ) [3,5]. As it was discussed in [3], these features can be ascribed to "forbidden" transitions $\Delta m = \pm 2, 3, 4...$ which may be allowed due to the mixing of states caused by non-diagonal terms in the spin Hamiltonian. Note that recently the similar mixing of states was considered as an essential factor affecting magnetic relaxation in NP's. [4]. A "giant spin" approach was also applied to the description of nuclear spin relaxation stimulated by nanoparticles used as contrast agents in MRI treatment [6].

However, in spite of a number of evidences for the quantal effects in NPs, many questions still remain open. The modeling of the EMR line shape proposed in Ref. [2], though qualitatively consistent with the experiment, includes a number of phenomenological fitting parameters. Temperature dependence of the low-field signal intensity demonstrates a sharp drop upon cooling which is too steep in comparison with the theoretical predictions [3]. The exact mechanism of these "forbidden" lines is not clear yet. According to the quantatization model [3], it can be related to the effects of magnetic anisotropy or interparticle dipole-dipole interactions. Thus, further study is necessary to clarify the origin of the effects and search for the border between classical and quantal phenomena in fine magnetic objects.

The idea of this work is to employ specially textured samples where randomly oriented nanoparticles are arranged in parallel chains, which are well aligned in one direction. Depending on the predominating mechanism, the model [3] predicts a specific angular dependence of the intensity of the double-quantum (2Q) transitions as compared with the anisotropy in the EMR spectrum measured on the aligned particles. The results obtained on such systems might clarify the origin of the effect as well as provide more independent data to confirm or reject the quantatization approach.

Well-arranged chains of the particles can be fabricated using a porous anodic alumina membrane (PAA). PAAs are commonly used in fabrication and exploring parallel magnetic nanowires. In particular, detailed data on magnetic anisotropy of Ni nanowires were presented in Refs. [7-13], including the interplay between the easy axis and easy plane behavior depending on the pore diameter. Thorough static magnetic measurements on iron oxide NPs placed into the channels were reported in Ref. [14]. However, the anisotropic EMR spectra in the later system, not to mention the multiple quantum effects, were not studied yet until now.

**2. Experimental**

In the experiment we used two different series (A and B) of $Fe_3O_4$ nanoparticles of nominally the same size but different magnetization. The chloroform suspensions of the particles with the diameter of 10±2.5 nm were purchased from Ocean Nanotech Inc. According to thermal gravitometric analysis, the solid content consists of 40 percent of iron oxide and 60 % of surfactant (oleic acid). Magnetization measurements were performed by SQUID magnetometer at room temperature in the field range 0 - 40 kOe. Magnetization curves for the both samples are consistent with the Langevin law; however the saturation value of the magnetization, M, was different for the different series, which can be related to the different quality of the samples. For the sample A, $M_s^A = 250 \pm 20$ Oe which is lower but relatively close to the bulk value of 450 Oe. Such decreased values are typical for NPs and can be related to surface effects [15-17]. Magnetization of the sample B, $M_s^B = 60 \pm 6$ Oe, was significantly lower, which can be ascribed to the poorer quality of the magnetic content, presence of multiple defects or different phases.

To prepare the textured samples, a porous alumina membrane with the dimensions 1cm x 1cm x 51µm, pore diameter of 35 nm and porosity of 15% purchased from Synkera Technologies was used as a template. Magnetic particles were embedded into the porous space of membrane by the standard technique of nanofiltration. As illustrated in Fig.1 (see also the data presented in Ref. [14]), the nanoparticles entered channels, forming parallel stacks.

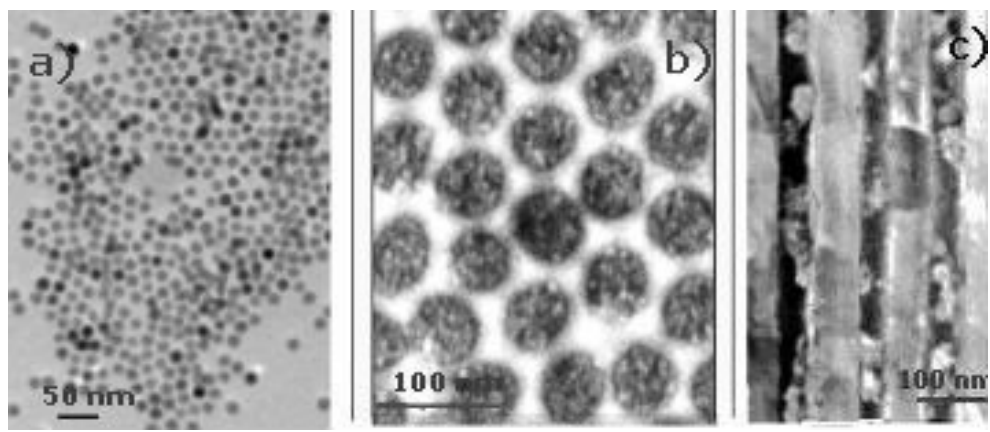

Fig. 1. a) TEM of the nanoparticles; b) and c) FESEM of the membrane channels with nanoparticles: b) view from the top; c) view from the edge

For the comparison purposes, samples with an arbitrary arrangement of the particles in solid polymer matrices were fabricated as well, following the method described in Ref. [5]. Particles were diluted in polystyrene in different concentrations, producing solid suspensions with 0.1 wt % and 0.01 wt % of the nanoparticles.

The experiments were performed using a standard Bruker EMX spectrometer, the microwave frequency was about 10 GHz (X band). The spectra were recorded using the standard field modulation technique, with the modulation amplitude 1 G and frequency of 100 kHz, resulting in the derivative of the EMR absorption. Angular dependences of EMR spectra were studied at room temperature. The orientation of the external magnetic field **B** with respect to the sample was adjusted through rotation of the sample in the horizontal plane, keeping the microwave field $B_1$ perpendicular to the external field **B**. A commercial gas flow cryostat was used in variable temperature experiments.

The EMR technique was also used for independent estimation of the NPs magnetization. The EMR intensity (the double integrated signal) was compared with that of paramagnetic reference ($CuSO_4 \cdot 5H_2O$). Within the experimental error (±10%), the results were consistent with the static measurements, indicating that nearly all magnetic material contributes to the EMR spectrum.

## 3. Results

The EMR spectra in the samples with the random orientation of NPs are presented in Fig.2 (for the convenience of comparison, the signals are normalized to the same integral intensity). The main EMR signal with the g-factor, g ≈ 2, consists of a broad line superimposed with a narrow one. Such a shape is typical for NPs as well as superparamagnetic and exchange-coupled clusters embedded in a non-magnetic matrix (see, for example, Refs. [2,5,18-22]). In Ref. [2], the spectrum was interpreted in the framework of the quantatization model. Additionally, a weak low-field signal is seen at g ≈ 4; this line was previously reported and assigned to double quantum (2Q) transitions in Refs.[3,5].

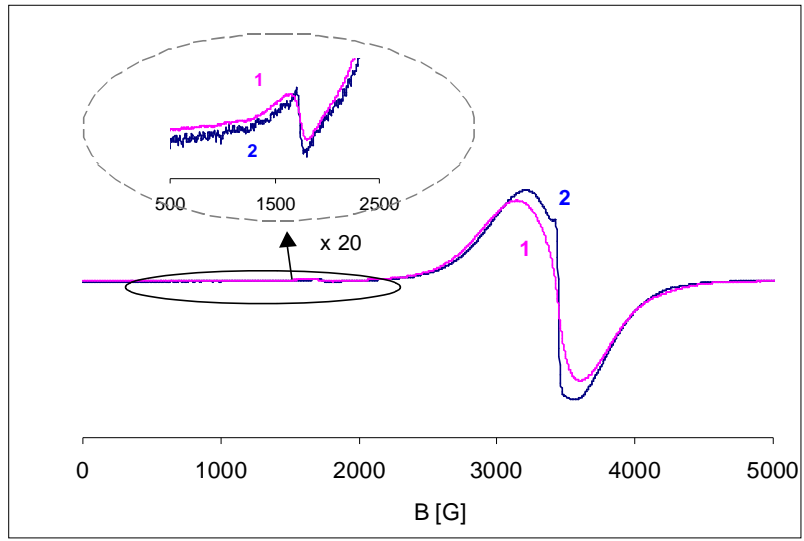

Fig. 2. EMR of NPs in polymer matrix with particle concentrations of 0.1 wt % (1) and 0.01 wt% (2). Inset: the double quantum transitions.

To distinguish this half-field resonance from a well-known EPR feature with g = 4.3 expected for paramagnetic $Fe^{3+}$ ions [23], the temperature dependence of the low-field spectrum is shown in Fig. 3. A rapid decrease of the low-field signal upon cooling unambiguously testifies against paramagnetic origin and in favor of 2Q identification (see Refs. [3,5] for more details).

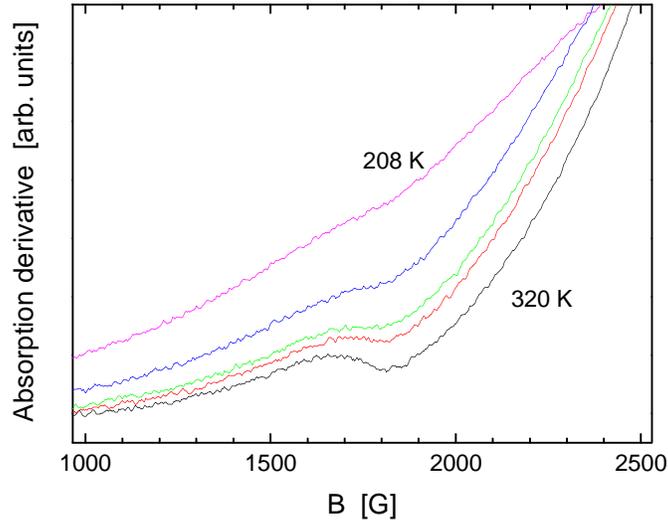

Fig. 3. The low-field feature in the EMR spectrum recorded at various temperatures. From the bottom to top: 320 K; 285 K; 273 K; 250 K; 208 K.

As is expected, the main EMR signal is broader in the sample with the higher concentration of NPs. In this sample, the sharp central feature is not well-resolved due to the broadening effect of inter-particle dipole-dipole interactions. The broadening of the 2Q signal is observed as well.

Let us now discuss the results obtained in the textured samples. The Fig. 4 demonstrates the EMR signals observed in the Sample A at various angles θ between the external magnetic field **B** and the direction of the channels.

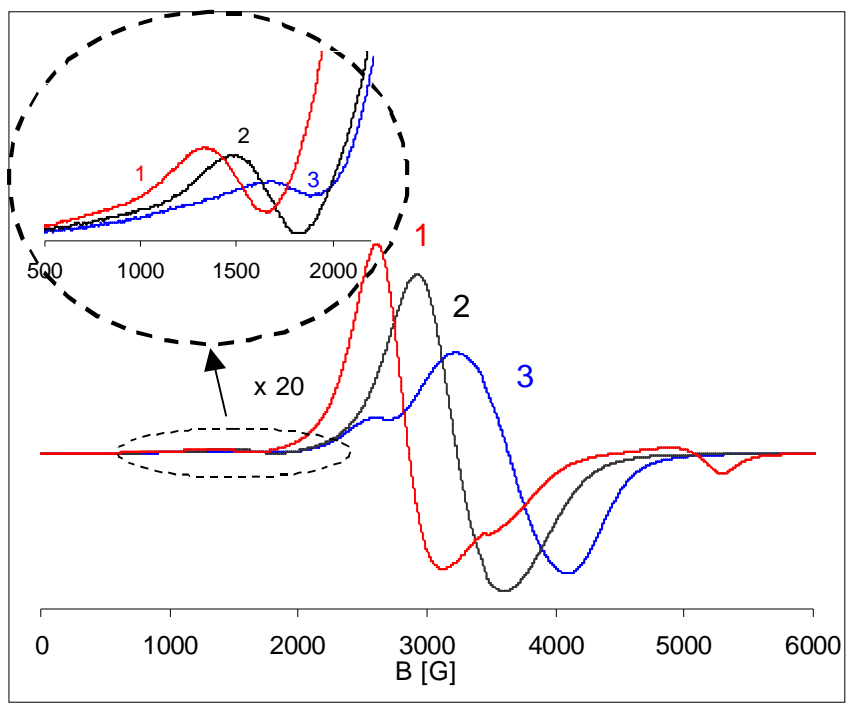

Fig. 4. a) EMR in the textured sample A for different orientations between the magnetic field and the long direction of the channels. $\theta = 2$ degree (Trace 1), 48 degree (Trace 2) and 87 degree (Trace 3).

The main EMR spectrum is broad, similar to that observed in the concentrated non-aligned sample (Fig. 2), pointing to a relatively close packing of the NPs in the matrix. The position of the main resonance strongly depends on the sample orientation, varying from lower fields at $\theta = 0$ (when the membrane channels are parallel to the external magnetic field) to higher fields at the perpendicular orientation. An additional weak and broad signal is observed as well with an opposite angular behavior (seen in Fig. 4 at ~ 5.2 kG, Trace 1, and 2.6 kG, Trace 3). It can be resulted from the particles distributed on the outer plane of the membrane or densely packed at the channel entrances, and is beyond the consideration of this paper. The position of the low-field signal depends on the sample orientation as well, see inset in Fig. 4.

The angular dependences of the peak positions of the main and 2Q signals, $B_r(1Q)$ and $B_r(2Q)$, are shown in Fig. 5. The both resonances reveal similar angular dependences typical of the uniaxial anisotropy. This result indicates a strong alignment of the particles in the membrane channels and clearly demonstrates that the main and half-field lines belong to the same entities.

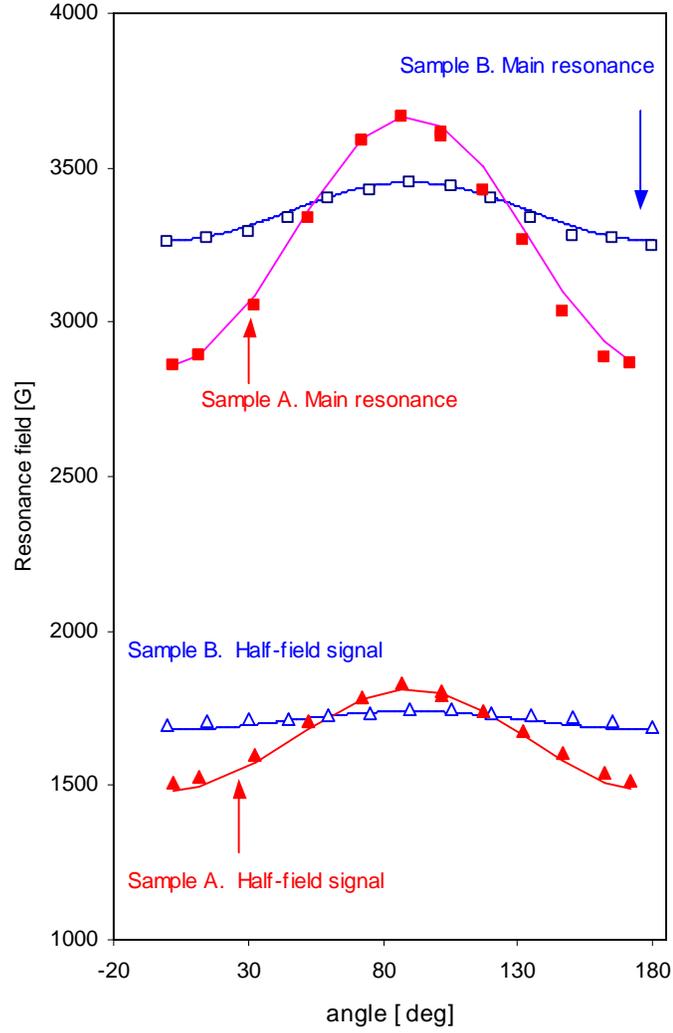

Fig. 5. Angular dependence of the resonance positions of the main resonance (cubes), and 2Q signal (triangles) in the textured samples A (filled symbols) and B (open). Solid lines are the Eqs. (1a), (1b).

The solid lines in Fig. 5 represent the fitting of the data as

$$B_r(1Q) = B_0 - bP_2(\theta) \qquad (1a)$$
$$B_r(2Q) = B_0/2 - b'P_2(\theta) \qquad (1b)$$

where

$$P_2(\theta) = (3\cos^2\theta - 1)/2$$

with the following parameters b, b′:

$$b^A = 530 \text{ G}; (b')^A = 200 \text{ G};$$
$$b^B = 120 \text{ G}; (b')^B = 50 \text{ G}$$

with the accuracy of about ±10%.

The angular dependence of the intensity of the half-field lines in the textured samples is one of the most important results of our experiments. As one can see in Fig. 4, the magnitudes of 2Q signals do significantly depend on the orientation of the external magnetic field. For the further analysis, the 2Q signals were extracted from the total spectrum (see examples on Fig. 6) and double integrated to estimate the intensity $I_{2Q}$. The ratio $I_{2Q}/I_{1Q}$ was plotted *vs* θ in Fig. 7, where the data for the non-textured samples (the dashed and dash-dotted lines) were shown for comparison. Note that the error bars in this figure are rather large and primarily caused by uncertainty in extracting the 2Q signals from the background formed by the wing of the main spectrum.

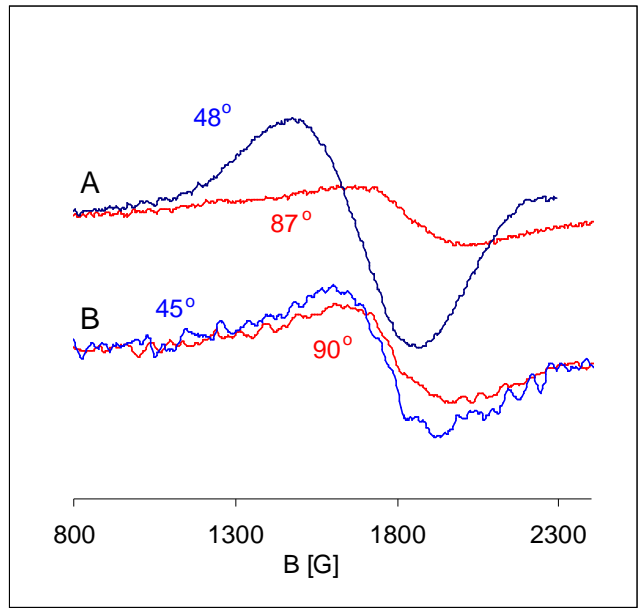

Fig. 6. 2Q signals extracted from the EMR spectrum in the textured samples A (upper traces) and B (lower traces) at various orientations of magnetic field (the orientations, θ, are indicated).

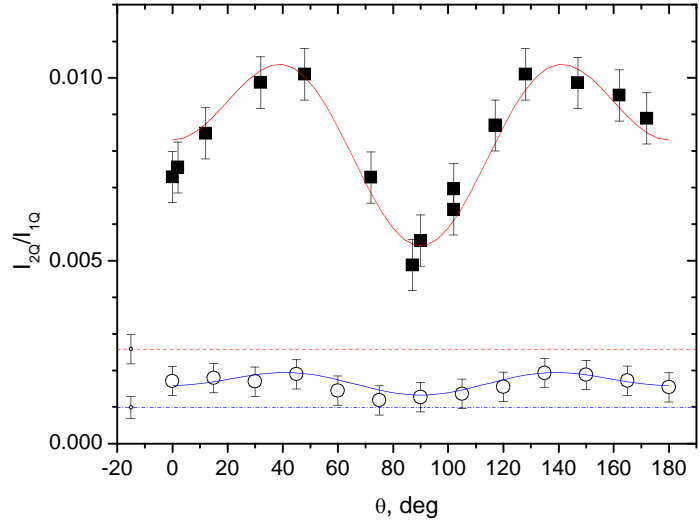

Fig. 7. Relative intensity of the 2Q line in the textured samples A (squares) and B (circles) in the dependence on the angle between the magnetic field **B** and direction of channels. The data for the non-aligned diluted suspensions of A and B are shown with the dashed and dash-dotted lines, respectively. The solid lines are calculated according to Eq.(6) with the parameters given in the text.

As one can see, the relative intensity of 2Q signals, $I_{2Q}/I_{1Q}$, in the both textured samples reveals minima at θ=0 and 90 deg. and a maximum at 45 deg. Such a dependence differs strongly from the angular dependence of the line position, see Eq. (1) and Fig. 5. It should also be noted that the 2Q intensity is considerably higher in the Sample A as compared to B, and in the textured samples as compared to the non-textured solid suspensions.

The detailed analysis is presented in the next Section. Note that the sample A with the relatively high magnetization demonstrates much greater anisotropy and 2Q effects than the sample B. Likely, the latter has a much broader distribution of the particle magnetic moments, that can lead to higher experimental errors and uncertainty in the interpretation. Therefore for the quantitative analysis we will use the data obtained in the sample A, whereas the sample B is used for a qualitative comparison.

## 4. Discussion

Let us first discuss the anisotropic shifts of the main EMR spectrum, see Figs. 4, 5 and Eq.(1a). These data unambiguously show that the particles inside the membrane channels are strongly aligned along the channel direction. Note that similar behavior was observed previously on liquid suspensions of NPs frozen in high magnetic fields [24, 2]; the effect was attributed to aligned aggregates (chains) formed by joint action of the external field and inter-particle dipolar interaction [2]. Since a layer of surfactant prevent the exchange interaction between adjacent particles, only the inter-particle dipole-dipole interaction should be taken into account. Note that our consideration does not require a compact filling of the channels with the particles. Instead, separate elongated aggregates are expected, randomly distributed along the channels. The magnetic interaction between the aggregates may contribute to the observed broadening of the EMR spectrum.

As the first approximation, we consider an aggregate as a strongly elongated ellipsoid filled by randomly distributed magnetic entities (spins) with an average magnetization $M_{av}$. In this case, the shift (first moment) of the EMR line reads [25]:

$$\Delta B_{el} = -(2/3) \cdot 2\pi M_{av} P_2(\theta) \qquad (2)$$

The factor 2/3 accounts for the static (z-z) part of the dipolar interaction only since the contributions from resonant (flip-flop) inter-particle interactions are suppressed by the inhomogeneity of the EMR line due to a random distribution of individual anisotropy axes. Considering the EMR line width, the single-particle anisotropy field (due to both the shape and crystalline sources) does not exceed few hundred Gauss, in agreement with the literature data [14].

It is worth noting that the factor 2/3 in Eq.(2) increases to unity for a ferromagnet with an easy axis directed along the channel; in such a case, in the terms of classical magnetism, Eq.(2) describes the effect of demagnetizing field. This approach was successfully used in studying magnetic nanowires in membranes [7-13].

Further, we suppose

$$M_{av} = c M_s L(\xi_0) \qquad (3)$$

Here $c$ is the concentration of pure magnetite in the aggregates and $L(\xi_0)$ is the Langevin function with $\xi_0 = M_s V B_0 / k_B T$, where V is the particle volume, T is the temperature, and $k_B$ is the Boltzmann constant. In our experiment, $M_s^A = 250$ Oe, $B_0 = 3.4$ kG, and T=295 K, which yield $L(\xi_0) = 0.9$. Then, substituting experimental $b^A = 530$ G for $\Delta B_{el}$ in Eq.(2) one gets $c = 0.56$. This value looks quite reasonable and suggests that nearly spherical NPs are packed closely enough in the elongated stacks inside the channels.

Let us analyze the data using another approximation as well. Consider a linear chain which is parallel to the channel axis and consists of spherical NPs with the diameter d and magnetic moment $\mu = M_s V$. Using the fact that the field outside of a uniform magnetic sphere is equivalent to the field of a point magnetic moment **μ** placed at the center of the sphere, the total z-component of the local field seen by every particle in a long linear chain is equal to:

$$\Delta B_{ch}(\theta) = 4.8 MV\, P_2(\theta)/d^3 = 0.8\pi M_s L(\xi_0) P_2(\theta) \qquad (4)$$

where the distance between adjacent particles is taken equal to d, $V = \pi d^3/6$, and the numerical factor is caused by summation over the chain. Comparing Eqs.(4) and (1a), one gets

$$b^A{}_{ch} = 0.8\pi M_s L(\xi_0) = 565\ G$$

in good agreement with the experimental value of $b^A = (530\pm 50)$ G. Thus, both approximations yield nearly the same result, confirming our basic assumptions described above.

Similar estimations for the Sample B lead to the satisfactory results as well. The shifts of the main EMR line are nearly proportional to the corresponding magnetizations M, see Eq.(1a). This is well expected in the case when observed anisotropic shifts of the EMR spectrum are caused mainly by the magnetic dipolar interactions between the NPs closely packed into stacks elongated predominantly in one direction.

Note that the columns formed by the NPs are likely not strictly parallel to each other simply because of considerable difference between the particle size and pore diameter. Instead, a distribution of the chain directions around the channel axis is expected resulting in some asymmetry in the observed line shape. This effect is indeed observable: as seen in Fig.4, the signals taken at the extremal orientations are asymmetric with their centers-of-gravity being shifted with respect to the points of zero derivative toward higher fields at θ~0 and lower fields at θ~π/2. In this case the center-of-gravity corresponds to the position averaged over the distribution of the anisotropy directions, whereas the absorption peaks (zero derivatives) are situated closer to those in strict parallel or perpendicular orientations. Simple estimations (not presented for brevity) show that the observed asymmetry can be explained with the Gaussian distribution of the chain directions with the normal deviation of 15-20 deg. This estimate will be used further.

Now, let us discuss the low-field (2Q) signals. In the frames of the quantatization model [3], the 2Q transitions are explained by the admixing of the quantum states with m, m±1. The two major possible mechanisms considered in [3] are: a) the magnetic anisotropy of an individual particle, which allows quantum transitions with Δm=2, or b) interactions between nanoparticles, when two particles coupled together through the dipole-dipole interaction absorb one photon of the microwave radiation. For the both mechanisms, the position of 2Q signal is expected to demonstrate the angular behavior similar to the behavior of the main resonance and be described with same Eqs.(2)-(4) but corrected for the reduced value of the magnetic field B. The decreased magnitude of the anisotropic shift of the half-field signals (Fig. 5) can be attributed to the lower M ∝ L(ξ) in the low field range. An additional apparent decrease in the effective anisotropy of the 2Q signal can also be related to the specific angular dependence of the 2Q-line intensity, which has minima in the both extremal orientations, θ=0 and π/2 (see below).

Let us consider the angular dependence of the relative intensity of the half-field resonance, Fig.7, and compare the data with the predictions of the quantum model. According to the theory, the intensity of the "forbidden" 2Q transitions to be observed in the field $B=B_0/2$ reads [3]:

$$\left(\frac{I_{2Q}}{I_{1Q}}\right)_{anis} = 8\sin^2\theta\cos^2\theta \left(\frac{B_a}{B_0}\right)^2 \frac{\left[\xi - 3L(\xi)\right]}{\xi^2 L(2\xi)} \equiv a_{anis}\sin^2\theta\cos^2\theta \qquad (5)$$

$$\left(\frac{I_{2Q}}{I_{1Q}}\right)_{dip} = 18\sin^2\theta_{ij}\cos^2\theta_{ij}\left(\frac{\mu}{Br_{ij}^3}\right)^2\frac{[L(\xi)]^2}{\xi L(2\xi)} \equiv a_{dip}\sin^2\theta_{ij}\cos^2\theta_{ij} \qquad (6)$$

Here B and $\xi$ correspond to the conditions of the half-field resonance. These formulas were derived in the first approximation of perturbation theory as applied to the spin Hamiltonian of the ground spin multiplet of a particle considered as a giant exchange coupled cluster. Eq.(5) corresponds to the case when the mixing of the states (which allows the 2Q transitions) is due to the single-particle magnetic anisotropy; as mentioned above, the individual anisotropy axes are directed randomly and cannot be responsible for the angular dependence. Eq.(6) accounts for the dipole-dipole interaction between particles i and j, where $r_{ij}$ and $\theta_{ij}$ are the polar coordinates of the radius-vector connecting the particles. In the closely packed columns elongated along the channel direction, one can expect $\theta_{ij} \approx \theta$ and $r_{ij} \approx d$.

The observed angular dependence of the 2Q intensity (Fig. 7) well resembles the $\sin^2\theta\cos^2\theta$ function which enters Eq.(6). Certainly, this finding qualitatively confirms the model and clearly indicates the major role of the interparticle mechanism. However, there are two "anomalous" features to be explained. First, the data in Fig.7 are lifted above the abscissa axis, pointing to a noticeable isotropic contribution. The second peculiarity is an asymmetry of the observed angular dependence: the minimum at $\theta=0$ is not as deep as that at $\theta=90$ deg. This anomaly can be explained by taking into account a spread of the directions of NP stacks in respect to the channel axis. Such a possibility was already discussed in connection with the asymmetry of line shape seen in Fig. 4. Taking into account that the channel diameter (35 nm) is considerably larger than the mean particle size (~10 nm), this suggestion looks quite reasonable.

Taking into account the distribution for the angles $\psi$ between the channel and chain axes, one gets the corrected angular dependence in the form:

$$\frac{I_{2Q}}{I_{1Q}} = \alpha\left\langle Sin^2\theta Cos^2\theta\right\rangle_\psi + \beta \qquad (7)$$

where the brackets denote averaging over the distribution and β stands for the isotropic contribution. Using the geometrical relation between θ and ψ and supposing the Gaussian distribution g(ψ) with a normal deviation $\psi_0$, we fitted the experimental data with the solid curves shown in Fig. 7. The agreement can be considered as satisfactory. In particular, the difference between the minima at perpendicular and parallel orientations is well reproduced by the fitting curves. The parameters found from the fitting are as follows:

$$\alpha^A=0.04;\ \beta^A=0.002;\ \psi_0^A=20^0;\ \alpha^B=0.004;\ \beta^B=0.001;\ \psi_0^B=17^0 \qquad (8)$$

The fitting errors are about ±20% for the sample A and considerably worse (about 50%) for B. In the frames of this accuracy, the results look quite reasonable. A larger value of the factor $\alpha^A$ as compared with $\alpha^B$ can be readily understood as a consequence of the difference in corresponding magnetizations which determine the dipolar strength. Further, the isotropic terms $\beta^{A,B}$ are close to the intensities obtained experimentally on the non-aligned samples, see the dashed lines in Fig. 7. Finally, the spread angles $\psi_0^{A,B}$ practically coincide with that obtained above from the analysis of the main spectrum.

Now let us compare the obtained $\alpha^{A,B}$ with the theoretical predictions of Eq. (6). Again, we will apply the linear chain/stack model used above in the interpretation of the anisotropic shifts. Substituting $r_{ij}=d$, $\theta_{ij}=\theta$, and performing summation over the chain (practically, only two nearest neighbors should be taken into account), one gets $a_{dip}^A = 0.032$ and $a_{dip}^B = 0.0023$, to be compared with the best-fit values of $\alpha^{A,B}$, Eq.(8). Taking into account the experimental errors and approximations accepted in the theoretical model, the agreement is good enough, especially for the sample A.

The isotropic contribution, β, to the 2Q intensity can be attributed, at least partially, to the single-particle anisotropy averaged over random orientations of the anisotropy axes. Using Eq. (5) and employing $\beta^{A,B}$ from Eq.(8), the anisotropy field $B_a$ of the order of $10^2$ G is estimated for the both samples. This is consistent with the observed width of the main resonance, suggesting the inhomogeneous broadening caused by magnetic anisotropy. An additional source of the isotropic 2Q signal may be related to the particles situated near the central axis of the channel and surrounded by the neighbors on all sides.

To get an overall picture illustrating the correlation between the model predictions and experimental data, let us compare the values of the normalized dipolar strength, $h_d \equiv \mu/d^3$, estimated from different points of view. First, values of $h_d$ for A and B samples are calculated using the magnetization data, Section 2. Second, these values are obtained from the shifts of the main resonance, by applying the linear chain model. Finally, the measured intensities of the 2Q signals are used, by equating the best-fit ($\alpha_{dip}$) and theoretical ($a_{dip}$) factors. The results are presented in Table I.

Table I.

Comparison of the dipolar strength values obtained by different methods

| The method | Magnetization curve | Anisotropic shift (linear chain model) | 2Q intensity |
| --- | --- | --- | --- |
| $h_d^A$, Oe | 131 ± 10 | 123 ± 10 | 140 ± 30 |
| $h_d^B$, Oe | 31 ± 6 | 43 ± 10 | 40 ± 20 |

One can see that within the experimental errors, the values of $h_d$ determined from independent measurements are consistent, confirming validity of the model.

In conclusion, elongated stacks of the magnetite nanoparticles have been fabricated by means of filtering the NP suspension through the porous alumina membrane. The positions of both the main EMR and weak half-field lines caused by "forbidden" 2Q transitions reveal the anisotropy typical of z-z dipolar interaction within the aligned columns, whereas the relative intensity of the 2Q signal demonstrates the $\sin^2\theta\cos^2\theta$ dependence predicted by the quantatization model [3]. The values of the normalized dipolar parameter, $h_d = \mu/d^3$, determined from independent experimental methods (the magnetization curves, anisotropic EMR shifts, and 2Q intensity) and theoretical calculations, are well consistent to each other for the both samples under study. Thus, good agreement with the experimental data is demonstrated, evidencing applicability of the quantum description to magnetic resonance and magnetization dynamics in NPs.

**Acknowledgments**

Authors would like to thank A. Pradhan for the magnetization measurements. The work was supported by National Science Foundation (NSF) PREM Grant # DMR-0611430, Russian Foundation for Basic Research (Grant 08-02-00040), and Russian Academy of Sciences (Program P-03).
**References**

1. J.L. Garcia-Palacios, S. Dattagupta, Phys. Rev. Lett. **95**, 190401 (2005).
2. N. Noginova, F. Chen, T. Weaver, E.P. Giannelis, A.B. Bourlinos, V.A. Atsarkin, J. Phys.: Condens. Matter **19**, 246208 (2007).
3. N. Noginova, T. Weaver, E.P. Giannelis, A.B. Bourlinos, V.A. Atsarkin, V.V. Demidov, Phys. Rev. B **77**, 014403 (2008).
4. D.A. Garanin, Phys. Rev. B **78**, 144413 (2008).
5. Maxim M. Noginov, N. Noginova, O. Amponsah, R. Bah, R. Rakhimov, V.A. Atsarkin, J. Magn. Magn. Mater. **320**, 2228 (2008).
6. A. Roch, R.N. Muller, J. Chem. Phys. **110**, 5403 (1999).
7. P.M. Paulus, F. Luis, M. Kröll, G. Schmid, L.G. de Jongh, J. Magn. Magn. Mater. **224**, 180 (2001).
8. A. Encinas-Oropesa, M. Demand, L. Piraux, I. Huynen, U. Ebels, Phys. Rev. B **63**, 104415 (2001).
9. U. Ebels, J.-L. Duvail, P.E. Wigen, L. Piraux, L.D. Buda, K. Ounadjela, Phys. Rev. B **64**, 144421 (2001).
10. M. Kröll, w.j. Blau, D. Grandjean, R.E. Benfield, F. Luis, P.M. Paulis, L.J. de Jongh, J. Magn. Magn. Mater. **249**, 241 (2002).
11. C.A. Ramos, M. Vazques, K. Nielsch, K. Pirota, J. Rivas, R.B. Wehrspohn, M. Tovar, R.D. Sanchez, U. Gösele, J. Magn. Magn. Mater. **272-276**, 1652 (2004).
12. C.A. Ramos, E. Vassalo Brigneti, D. Navas, K. Pirota, M. Vázques, Physica B: Condens. Matter **384**, 19 (2006).
13. E.P. Hernández, S.M. Rezende, A. Azevedo, J. Appl. Phys. **103**, 07D506 (2008).
14. M. Pasquale, E.S. Olivetti, M. Coïsson, P. Rizzi, G. Bertotti, J. Appl. Phys. **103**, 07D527 (2008).

15. A.A. Novakova, V.Yu. Lanchinskaya, A.V.Volkov, T.S. Gendler, T.Yu. Kiseleva, M.A. Moskvina, S.B. Zezin. J. Magn. Magn. Mater. **258-259**, 354 (2003).
16. T. Kim, M. Shima, J. Appl. Phys. **101**, 09M516 (2007).
17. P. Dutta, S. Pal, M.S. Seehra, N. Shah, G.P. Huffman, J. Appl. Phys. 105, 07B501 (2009).
18. V.K Sharma, F. Waldner, J. Appl. Phys. **48**, 4298 (1977).
19. F. Gazeau, V. Shilov, J.C. Bacri, E. Dubois, F. Gendron, R. Perzynski, Yu.L. Raikher, V.I. Stepanov, J. Magn. Magn. Mater. **202**, 535 (1999).
20. D. Prodan, V.V. Grecu, M.N. Grecu, E. Tronc, J.P. Jolivet, Meas. Sci. Technol. **10**, L41 (1999).
21. R. Berger, J. Kliava, J.-C. Bissey, V. Baietto, J. Appl. Phys. 87, 7389 (2000).
22. N. Noginova, R. Bah, D. Bitok, V.A. Atsarkin, V.V. Demidov, S.V. Gudenko, J. Phys.: Condens. Matter **17**, 1259 (2005).
23. D. J. E. Ingram, *Biological and Biochemical Applications of Electron Spin Resonance*, Adam Hilger LTD, London, 1969, Chapt. 6.7.
24. F. Gazeau, J.C. Bacri, F. Gendron, R. Perzynski, Yu.L. Raikher, V.I. Stepanov, E. Dubois, J. Magn. Magn. Mater. **186**, 175 (1998).
25. A. Abragam and M. Goldman, *Nuclear Magnetism: Order and Disorder*, Clarendon Press, Oxford, 1982, Chapt. 5.